\documentstyle[11pt,newpasp,epsf]{article}
\def\lax    {${_<\atop^{\sim}}$}

\def\msol     {{M$_{\odot}$}}
\def\ergs   {~erg~s$^{-1}$}
\def\etal   {{et~al}}

\def\edcomment#1{\iffalse\marginpar{\raggedright\sl#1\/}\else\relax\fi}
\reversemarginpar

\begin{document}
\title{Chandra Observations of M31}
\author{M.R. Garcia, S.S. Murray, F.A. Primini, W.R. Forman, C. Jones,
J.E. McClintock}
\affil{Smithsonian Astrophysical Observatory, 60 Garden St., Cambridge
MA, 02138, USA}

\begin{abstract}

	As part of the Chandra GTO program we are monitoring and
surveying M31 using the HRC and ACIS cameras.  These observations
have resolved the nuclear X-ray source into five separate sources, one
of which is very soft and may (or may not!) be associated with the central
super-massive black hole.  In addition, the superb spatial resolution
and low scattering of the Chandra telescope allows us to unambiguously
resolve the diffuse emission from the point sources.  This emission is
clearly softer than the point sources, and also increases with
temperature radially.   The monitoring nature of the observations
allows detailed study of the variability of the point sources. 

\end{abstract}

{\bf Introduction:} M31 has been imaged with Chandra monthly as part
of the AO1 GTO program.  Each month five HRC observations provide a
complete image of the galaxy to a sensitivity of $10^{37}$\ergs .
Transient sources are re-observed with the ACIS camera, typically
yielding $\sim 100$ counts per source.  Thus the known (Galactic)
types of accretion powered x-ray transients are being detected, their
spectra measured, and their lightcurves studied.  In the first
observation we discovered a bright transient $26''$ to the
west of the nucleus (Murray \etal\/ 1999), which may be associated
with a stellar mass black hole.  The next 8~ACIS images were centered
on the nucleus, to follow the evolution of this transient.
More recently a second bright transient was discovered in the HRC
images which included M32 (Garcia \etal\/ 2000a).  The most recent
ACIS imaging has therefore been centered on this transient, which is
likely in M32.

During AO2 the GTO monitoring program will be cut in $\sim 1/2$, but there
are two GO programs which also will image M31.  A program lead by
R.~Di~Stefano will use the ACIS to repeatedly image the area surrounding
three super-soft sources, and a program lead by P.~Kaaret will use the
HRC to study the variability of sources in the nuclear region.  
These studies (and others likely to come) should allow much of what has
been considered `galactic X-ray astronomy' to be carried out in our
nearest neighbor galaxy, M31. 


\begin{figure} 
\plotfiddle{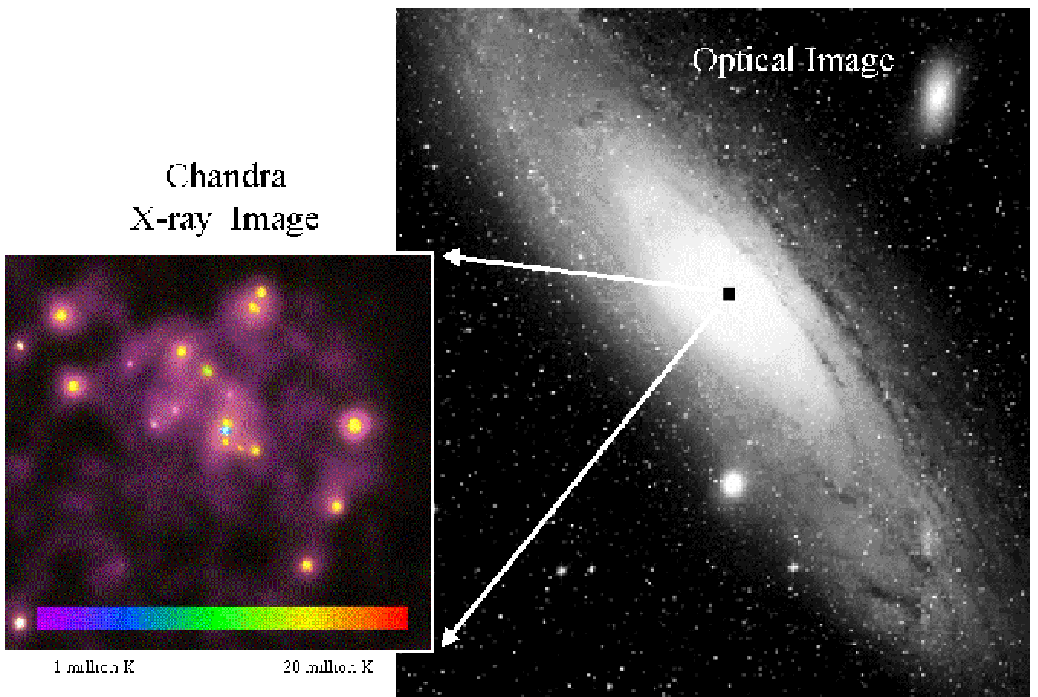}{2.75in}{0}{100}{100}{-144}{+12}
\vskip -0.1in
\begin{minipage}[t]{5truein}
Figure 1. Chandra and DSS images of M31. The
L-shaped central region contains 5~sources, one of which is very soft
(see Garcia \etal\/ 2000b).
\end{minipage}
\vskip -0.2in
\end{figure}

{\bf The Nucleus:}~ The central object seen with the ROSAT HRI is
clearly resolved into 5 sources by Chandra (Figure 1).  The celestial
location of these sources can be determined to $\sim 1''$ based on the
Chandra aspect solution alone.  To verify the Chandra coordinates we
registered the X-ray image against the Bologna catalog of M31 globular
clusters (Battistini \etal\/ 1987).  Based on the positional
coincidence of the soft nuclear source and the radio/UV nucleus, we
previously suggested that these sources may be associated (Garcia
\etal\/ 2000b).  However,
the positional accuracy was insufficient to exclude an association
with another source located $\sim 1''$ to the north
(CXO~J004244.2+411609; Garcia \etal\/ 2000b); furthermore, it is also
possible that neither source is associated with the nucleus.

By finding UV counterparts in HST images, it should be possible to
determine the position of the Chandra nuclear sources relative to the
UV nucleus to $\sim 0.1''$.  This is challenging given the small
region of M31 currently covered by HST images and the expected UV
faintness of likely counterparts.  Thus far we have been able to find
only a single secure common source (the globular cluster MITa 213,
Magnier 1993).  This allows us to determine the translational offset
between the X-ray and UV images, but does not allow us to determine
the rotational offset.  If we assume there is no rotational offset,
then the bright UV nucleus is within $\sim 0.5''$ of
CXO~J004244.2+411609, and is $\sim 1.5''$ North of the soft nuclear
source.  We caution that this registration is inconclusive because the
possible rotational offset could create $\sim 1''$ position errors.

The soft nuclear source is highly variable.  The emitted luminosity as
seen on 6 separate Chandra observations is shown in Figure~2.  While
previous X-ray missions were not able to resolve this source from its
surrounding 4 neighbors, those neighbors appear relatively constant in
the Chandra data.  Assuming they are constant allows us to determine
that the soft nuclear source had an emitted luminosity of $\sim 4
\times 10^{38}$ \ergs\/ when first observed with the Einstein
observatory (Van Speybroeck \etal\/ 1979).  The calculation of the
emitted luminosity is very sensitive to the assumed absorption, and
 the peak emitted luminosity may have been as low as $1.2
\times 10^{38}$ \ergs .

\begin{figure} 
\plotfiddle{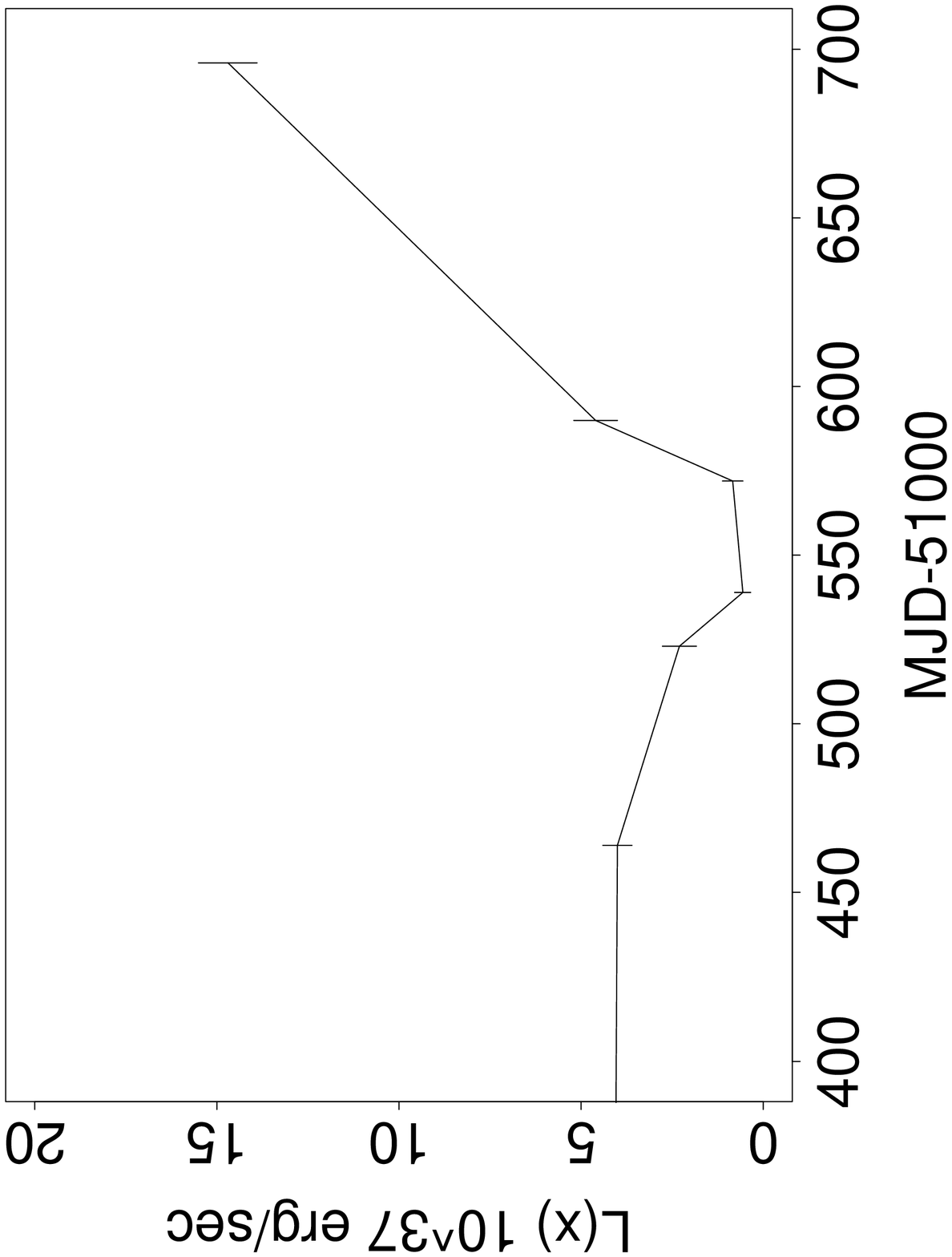}{1.75in}{-90}{25}{25}{-90}{140}
\vskip 0.1in
\begin{minipage}[t]{5truein}
\hskip 0.5in \hbox{Figure 2. Chandra Light Curve of the M31 Soft Nuclear Source}
\end{minipage}
\vskip -0.2in
\end{figure}

{\bf Transients:} In addition to the two bright transients mentioned above,
there are numerous fainter transients seen in the ACIS
images of the M31 nuclear region.  We define a `transient' as a
source that is clearly detected on one image and clearly absent on
another.  For the faintest of these transients, the range of
variability is constrained to be only a factor of $>3$, while for the
brightest it is a factor of $>30$.  Within the central $4'$ box there
are 8 transients out of a total population of 40 sources.  Previous
estimates for the number of transients in M31 were lower ($\sim 6$\%;
Primini, Forman and Jones 1993), Our current estimate of $\sim 20$\% transient
sources is comparable to the estimate for the transient population in
the Milky Way ($\sim 16$\%; Primini \etal\/ 1993).

The two brightest transients have peak luminosities \lax~$10^{38}$
\ergs , consistent with  X-ray novae seen in the Galaxy.  Two
somewhat fainter transients have peak luminosities $\sim 10^{37}$\ergs
, consistent with the pulsing X-ray transients and post-minimum period
transients (King 2000) seen in the Galaxy.  Five other transients have
peak luminosities of only $\sim 10^{36}$ \ergs .  While this is
consistent with pulsing X-ray transients in the Galaxy, one does not
expect to find large numbers of such objects in the bulge of M31: they
are associated with young stars, not the old stellar population of the bulge.
The recent discovery of an X-ray novae in the Galaxy with a peak outburst
luminosity of only $\sim 2 \times 10^{35}$ \ergs\/ (XTE J1118+480; 
McClintock \etal\/ 2000) emphasizes that these object do not always
reach peak luminosities near Eddington, and that therefore the fainter
transients in M31 could indeed be  X-ray novae.

{\bf Diffuse Emission:}  The azimuthally-averaged
surface brightness profile and hardness ratio of the M31 diffuse emission
is shown in Figure~3 of Primini \etal\/ (2000).  It is clear that the
diffuse emission is substantially softer than nearly all the
point sources. The hardness ratio within the inner $3'$ is 
that expected from a Raymond-Smith thermal spectrum with a
temperature kT$\sim 0.3$~keV, consistent with the temperature found by
XMM for this diffuse emission (Shirey \etal\/ 2000).  Beyond $3'$ the
temperature of the diffuse emission increases until it reaches a
hardness ratio consistent with that typical of the point sources.

{\bf Point Source Luminosities:} Figure~3 shows a histogram of the
point source luminosities, based on the first 8.8~ks observation of
M31.  The histogram is consistent with the
break in the luminosity function at $\sim 10^{37.5}$ seen
with ROSAT (Primini \etal\/ 1993) and XMM (Shirey \etal\/ 2000).
The luminosity function is incomplete below $\sim
10^{36}$\ergs , but the sum total GTO observations should extend the
detection threshold to \lax~$10^{35}$\ergs .
The brightest source has ${\rm L_x \sim 1.4 \times
10^{38}}$\ergs , consistent with the Eddington limit for a solar mass
star.  This is very different from what is seen in star-forming galaxies,
where the peak point source luminosity approaches $10^{40}$\ergs\/
(Prestwich 2000; Fabbiano 2000).  The most straightforward
explanation for these very high luminosities is that these sources are
$\sim 100$\msol\/ accretors (Kaaret \etal\/ 2000).  However, it is
relevant to note that the $\sim 10$\msol\/ black holes in our galaxy
have been seen to have peak X-ray luminosities which are
super-Eddington.  For example, the X-ray nova
SAX~J1819.3-2525 had a peak luminosity of $10^{39.5}$\ergs , likely
due to beaming of the X-ray flux in our direction (Orosz \etal\/
2000).  X-ray novae often have jets (Fender 2000) and
therefore may often have beamed X-ray emission.

\begin{figure}\label{lumhist}
\plotfiddle{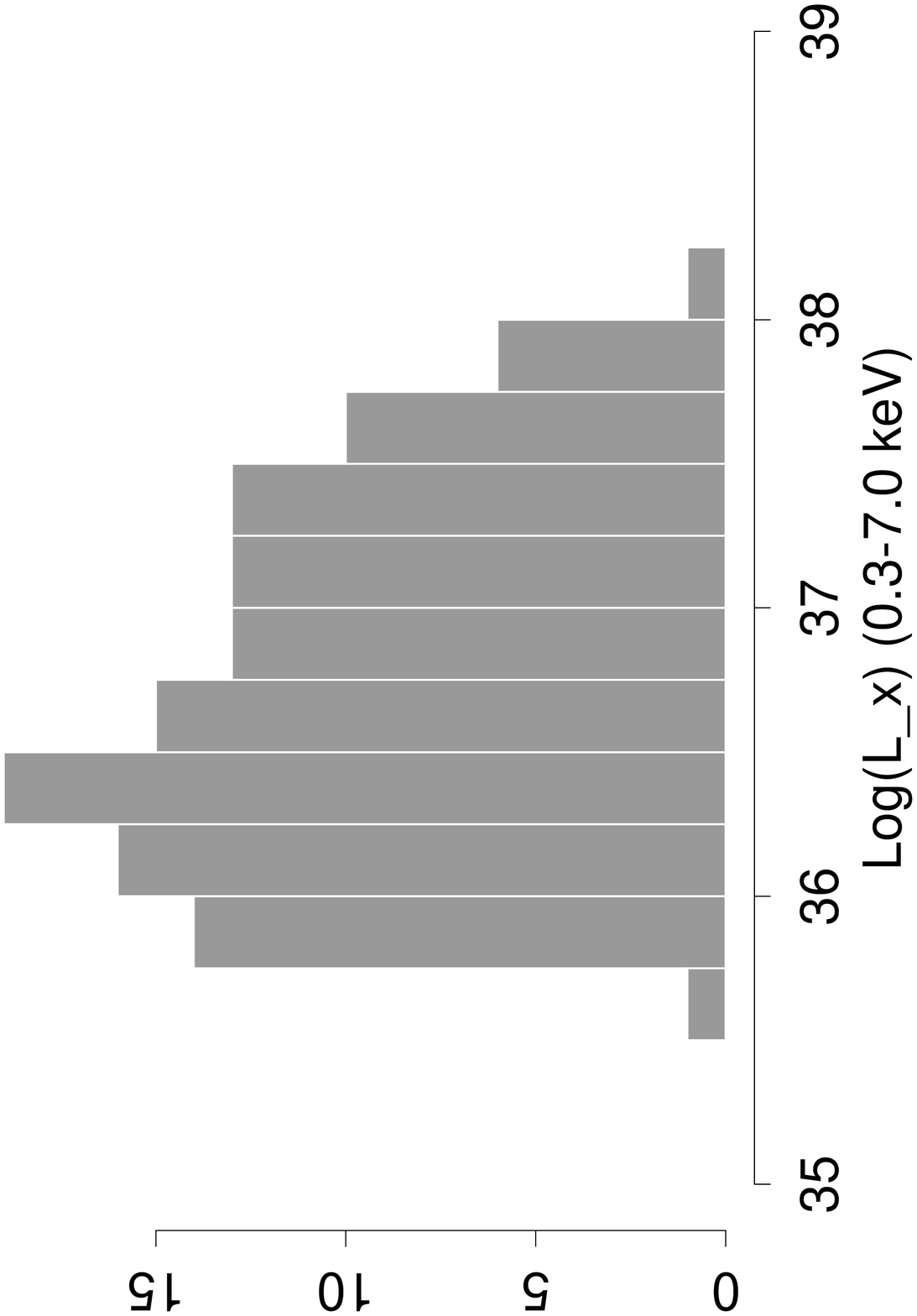}{2.00in}{-90}{30}{30}{-108}{180}
\vskip -0.05in
\begin{minipage}[t]{5truein}
\hskip 0.5in \hbox{Figure 3.  Histogram of M31 Point Source Luminosities}
\end{minipage}
\vskip -0.25in
\end{figure}


\vspace{-18pt}
\begin{references}
\vskip -10pt

\reference Battistini, P. L., Bonoli, F., Braccesi, A., Federici, L., Fusi
Pecci, F., Marano, B., \& Borngren, F. 1987, A\&AS, 67, 447

\reference Fabbiano, G.  2000, these proceedings

\reference Fender, R.P. 2000, MNRAS in press, astro-ph/0008447

\reference Garcia, M.R.,  Murray, S.S, Primini, F.A., McClintock,
J.E., Callanan, P.J. 2000a, IAUC 7498

\reference Garcia, M.R., Murray, S.S, Primini, F.A., Forman, W.R.,
 McClintock, J.E., \& Jones, C. 2000b ApJ 537, L23

\reference P. Kaaret, A.H. Prestwich, A. Zezas, S.S. Murray,
D.-W. Kim, R.E. Kilgard, E.M. Schlegel and M.J. Ward 2000, MNRAS in
press, astro-ph/0009211 

\reference King, A.R. 2000 MNRAS 315, L33

\reference Magnier, E.A., 1993 PhD Thesis, MIT.

\reference McClintock, J.E., \etal\/ 2000, submitted to ApJ (letters)

\reference Murray, S.S, Garcia, M.R, and  Primini, F.A. 1999, IAUC 7291 

\reference Orosz \etal\/ 2000, in prep for ApJ, also BAAS 32, 4, 83.20

\reference Prestwich, A., 2000, these proceedings

\reference Primini, F.A., Forman, W. and Jones. C. 1993 ApJ 410, 615

\reference Primini, F.A., Garcia, M.R, Murray, S.S., Forman, W.,
Jones, C., \& McClintock, J. 2000,"The Interstellar Medium in M31 and M33"
eds. E. Berkhuijsen \& R. Beck,	Shaker Verlag:Aachen, p. 145

\reference Shirey, R., 2000, A\&A in press, astro-ph/0011244

\reference van Speybroeck, L.; Epstein, A.; Forman, W.; Giacconi, R.;
 Jones, C.; Liller, W.; Smarr, L. ApJ 234, L45 


\end{references}
\end{document}